\pdfoutput=1
\documentclass[aps,prb,reprint,showkeys,floatfix]{revtex4-2}

\usepackage[T1]{fontenc}
\usepackage{graphicx}
\usepackage{upgreek}
\usepackage{color}
\usepackage{xspace}
\usepackage{textcomp}
\usepackage{siunitx}
\usepackage{amsmath}
\usepackage{amssymb}
\usepackage{physics}
\usepackage[colorlinks,
linkcolor=black,
citecolor=black,
filecolor=black,
urlcolor=blue,
pdfusetitle]{hyperref}
\hypersetup{
pdfauthor={Robert Bartel, Elias Lettl, Patrick Seiler,
 Thilo Kopp, German Hammerl},
pdfsubject={Solid-State Physics},
pdfkeywords={magnetotransport, spin–orbit coupling, oxide heterostructures,
2D, weak antilocalization, multiband, Hofstadter bands}
 }
\usepackage[noabbrev, capitalise]{cleveref}

\usepackage{lmodern}
\usepackage{libertinus}
\usepackage{libertinust1math}
\usepackage[section]{placeins}
\usepackage[english]{babel}
\usepackage[babel]{microtype}
\usepackage{mathtools}
\let\oldcite\cite
\renewcommand{\cite}{~\oldcite}
\makeatletter
\ifx\c@figure\undefined
\newcounter{figure}
\fi
\makeatother

\newcounter{subfigure}[figure]

\newcommand{\phantomsubcaption}{%
\addtocounter{figure}{1}%
\refstepcounter{subfigure}%
\addtocounter{figure}{-1}%
}

\numberwithin{equation}{section}

\Crefname{figure}{Figure}{Figure}
\Crefname{equation}{Equation}{Equation}
\Crefname{section}{Section}{Section}

\newcommand{\perovskite}[2]{\texorpdfstring{#1#2O$_3$}{#1#2O\_3}\xspace}
\newcommand{\BPO}{\perovskite{Ba}{Pb}}
\newcommand{\STO}{\perovskite{Sr}{Ti}}
\newcommand{\LAO}{\perovskite{La}{Al}}
\newcommand{\TO}{TiO$_2$\xspace}
\newcommand{\eg}{e.\,g.\xspace}
\newcommand{\ie}{i.\,e.\xspace}
\newcommand{\Tref}[1][]{\ensuremath{T_\mathrm{ref{#1}}}\xspace}
\newcommand{\tauso}[1][]{\ensuremath{\tau_\mathrm{so{#1}}}\xspace}
\newcommand{\tauo}[1][]{\ensuremath{\tau_\mathrm{e{#1}}}\xspace}
\newcommand{\tauphi}[1][]{\ensuremath{\tau_\phi}\xspace}
\newcommand{\Bso}[1][]{\ensuremath{B_\mathrm{so{#1}}}\xspace}
\newcommand{\Bo}[1][]{\ensuremath{B_\mathrm{e{#1}}}\xspace}
\newcommand{\Bphi}[1][]{\ensuremath{B_\phi}\xspace}
\newcommand{\Rs}{\ensuremath{R_\square}\xspace}
\newcommand{\RsT}{\ensuremath{R_\square(T)}\xspace}
\newcommand{\WAL}{\text{WAL}\xspace}
\newcommand{\QI}{\text{QI}\xspace}
\newcommand{\EEI}{\text{EEI}\xspace}
\newcommand{\MR}{\text{MR}\xspace}
\newcommand{\CWL}{\ensuremath{C^\mathrm{WL}}\xspace}
\newcommand{\CWAL}{\ensuremath{C^\mathrm{WAL}}\xspace}
\newcommand{\CEEI}{\ensuremath{C^\mathrm{EEI}}\xspace}

\newcommand{\kB}{\ensuremath{k_\text{B}}\xspace}
\newcommand{\muB}{\ensuremath{\mu_\text{B}}\xspace}

\newcommand{\tsampleone}{\SI{15.0}{\nano\meter}\xspace}
\newcommand{\Bsosampleone}{\SI{0.22}{\tesla}\xspace}   
\newcommand{\zetasampleone}{\SI{0.91}{}\xspace} 
\newcommand{\alphasampleone}{\SI{1.99}{}\xspace}
\newcommand{\betasampleone}{\SI{0.14}{\milli\tesla\per\raiseto{\alpha}\kelvin}\xspace}
\newcommand{\gammasampleone}{\SI{7.50}{\milli\tesla}\xspace}
\newcommand{\Trefsampleone}{\SI{11.1}{\kelvin}\xspace}

\newcommand{\Bsosampletwo}{\SI{0.13}{\tesla}\xspace} 
\newcommand{\zetasampletwo}{\SI{0.97}{}\xspace} 
\newcommand{\Bsosampletwoold}{\SI{0.10}{\tesla}\xspace} 

\newcommand{\Bsosamplethree}{\SI{0.24}{\tesla}\xspace}  
\newcommand{\zetasamplethree}{\SI{0.89}{}\xspace} 
\newcommand{\Bsosamplethreeold}{\SI{0.23}{\tesla}\xspace} 
\newcommand{\zetasamplethreeold}{\SI{0.84}{}\xspace}

\newcommand{\TwoD}{2D\xspace}

\DeclareMathOperator{\cc}{c}
\DeclareMathOperator{\T}{T}
\DeclareMathOperator{\Hopt}{H}
\DeclareMathOperator{\J}{J}
\newcommand{\ropt}{\ensuremath{\mathbf r}}
\newcommand{\Jopt}{\ensuremath{\mathbf J}}

\newcommand{\ci}{i}

\makeatletter
\newcommand{\int@BZ}[1]{%
\int_{\mathmakebox[\widthof{$\mathsurround=0pt #1 \int$}%
-\widthof{$\mathsurround=0pt #1 \int_{\rlap{}}$}][l]{%
\varOmega_{\mathrm{BZ}}%
}%
}%
}
\newcommand{\intBZ}{
\mathpalette\int@BZ\relax
}
\makeatother

\renewcommand{\vec}{\boldsymbol}

\let\originalleft\left
\let\originalright\right
\renewcommand{\left}{\mathopen{}\mathclose\bgroup\originalleft}
\renewcommand{\right}{\aftergroup\egroup\originalright}

\begin{document}

\title{Magnetotransport of Functional Oxide Heterostructures Affected by Spin–Orbit Coupling:
\texorpdfstring{\\}{}
A Tale of Two-Dimensional Systems}

\author{Robert Bartel}
\author{Elias Lettl}
\author{Patrick Seiler}
\author{Thilo Kopp}

\author{German Hammerl}
\email{german.hammerl@physik.uni-augsburg.de}
\affiliation{
Center for Electronic Correlations and Magnetism,
Experimental Physics VI,
Institute of Physics,
University of Augsburg,
86135 Augsburg,
Germany
}

\date{\today}


\begin{abstract}
Oxide heterostructures allow for detailed studies of 2D electronic
transport phenomena. Herein, different facets of magnetotransport in selected
spin–orbit-coupled systems are analyzed and characterized by their single-band
and multiband behavior, respectively. Experimentally, temperature- and magnetic
field-dependent measurements in the single-band system \BPO/\STO reveal strong
interplay of weak antilocalization (WAL) and electron–electron interaction
(EEI). Within a scheme which treats both, WAL and EEI, on an equal footing a
strong contribution of EEI at low temperatures is found which suggests the
emergence of a strongly correlated ground state. Furthermore, now considering
multiband effects as they appear, \eg, in the model system
\LAO/\STO, theoretical investigations predict a huge impact of filling on the
topological Hall effect in systems with intermingled bands. Already weak band
coupling produces striking deviations from the well-known Hall conductivity that
are explainable in a fully quantum mechanical treatment which builds upon
the hybridization of intersecting Hofstadter bands.
\end{abstract}

\maketitle

\section{Introduction}

Perovskite-related oxides show a huge variety of intrinsic properties.\cite{zubko:2011} With
oxide heterostructures, it is not only possible to combine such material
characteristics but also to identify novel electronic phases emerging on the
nanoscale which allows to trigger a plethora of functionalities.\cite{hwang:2012,Mannhart2010}
At the interfaces of certain polar insulators confined metallic electronic systems appear driven
by electronic reconstruction.\cite{hesper:2000,okamoto:2004} In addition, inversion symmetry is systemically
broken, a key ingredient for strong Rashba-type spin–orbit coupling, leading to
anomalous transport properties which will be addressed in this article. Moreover,
such electronic systems, when gapped, may assume
nontrivial values of topological invariants causing a particular
behavior of their magnetotransport. In fact,
magnetotransport allows to obtain a fingerprint of the electronic state of
metals, especially also of oxide heterostructures with their complex electronic
properties controlled by sizable spin–orbit coupling, multiband behavior, disorder, and Coulomb interaction.

This article covers two complementary spin–orbit-coupled electronic systems,
both with regard to magnetotransport: a disordered and a defect-free
2D system. Correspondingly, the article is organized as
follows:
In \cref{sec:disorder}, we examine experimentally \BPO thin-films grown
on \STO. The perovskite-related oxide \BPO is a single-band metal. In this
system with Rashba spin–orbit coupling disorder accounts for weak
antilocalization (WAL) in the presence of electron–electron interaction (EEI).
We briefly introduce these theoretical concepts of quantum corrections to
transport properties before we analyze our temperature- and magnetic field-dependent measurements. We then self-consistently extract parameters describing
spin–orbit coupling and EEI---indicating a correlated ground state in \BPO.
In a further step toward a general understanding it suggests itself to consider
the spin–orbit coupling dominated magnetotransport beyond the single-band \TwoD
systems.
In \cref{sec:theoretical-studies}, we analyze the influence of magnetic
fields on the transport properties of a defect-free \TwoD multiband system in
the fully quantum mechanical treatment of linear response theory. Our work is
inspired by the fact that magnetotransport studies of \LAO/\STO interfaces under
applied hydrostatic pressure can lead to counterintuitive results if evaluated
with standard semiclassical techniques.\cite{Seiler2018} However, as the
semiclassical Boltzmann transport theory builds upon a single-band model its
validity in case of multiband systems like \LAO/\STO should be
questioned.\cite{Kim2013} This is especially true if one expects topological
band aspects to play a fundamental role. After a general model description, we
start by analyzing magnetotransport for the single-band case revisiting the
results of the Hofstadter model. In a next step, we discuss multiband behavior
affected by atomic or Rashba-type spin–orbit coupling.

\section{Magnetotransport in Single-Band Systems Governed by Disorder} \label{sec:disorder}

Recently, we found that \BPO thin-films grown on (001)-oriented \STO single
crystals show single-band behavior and a pronounced magnetoresistance (MR)
which at low magnetic fields is evidently ruled by WAL.\cite{seiler:2019a}
Surprisingly, temperature-dependent measurements of the sheet resistance~\RsT account for an insulating low-temperature state, contradicting the
WAL result of magnetoconductance. Such a
counterintuitive behavior of thin-film samples
was observed before.\cite{liu:2011,wang:2011a} It is argued that \MR and \RsT
may originate from distinct sensitive channels leading to different
measurement-dependent ground
states.\cite{liu:2011,wang:2011a,amaladass:2017,lu:2011} By carefully
investigating \MR and \RsT, we unveiled that the expected WAL contribution in
\RsT is covered by a pronounced EEI contribution. However, up to now, we
neglected the mutual effect of EEI to MR as we considered it to be small.

Before we examine the influence of EEI on the WAL signal in our
samples, let us discuss the generic temperature and magnetic field dependencies on the
quantum corrections of the electrical transport of a disordered
\TwoD system.

Due to weak disorder low-temperature electronic transport in \TwoD materials is
affected by quantum interference (QI) resulting either in
insulating or metallic ground states. QI contributes significantly to the
electrical transport only if the electrons' temperature-dependent dephasing
time~\tauphi is large compared with, \eg, the elastic
scattering time~\tauo: randomly scattered electrons will unavoidably
self-interfere constructively with their time-reversal counterparts leading to
WL with its insulating ground
state.\cite{anderson:1958,abrahams:1979,hikami:1980,altshuler:1980a,lee:1985}
Pronounced spin–orbit (SO) coupling described by a timescale $\tauso$ associated
with the D'yakonov--Perel' spin relaxation ($\tauso\ll\tauphi$) instead contributes an
additional phase causing WAL which induces a metallic ground
state.\cite{hikami:1980,bergmann:1984,iordanskii:1994,dyakonov:1972}

Both QI effects, WL and WAL, are characteristically influenced by applied
time-reversal symmetry-breaking external magnetic fields which makes it possible to
experimentally decide on the type of quantum corrections. A comprehensive
description of the magnetic field-dependent first-order quantum correction to
the conductivity of an ideal \TwoD material is given by the
well-accepted Iordanskii--Lyanda-Geller--Pikus theory which relates the
specific magnetic field dependence to the winding number of the spin
expectation value around the Fermi
surface.\cite{iordanskii:1994,knap:1996,pikus:1995} In case of triple spin
winding, found in, \eg, \STO-based \TwoD thin-films,\cite{pai:2018,nakamura:2012,seiler:2018} the
Iordanskii--Lyanda-Geller--Pikus theory merges to the analytical result of the
Hikami--Nagaoka--Larkin theory.\cite{hikami:1980} The first-order quantum
correction to the conductivity~$\sigma$ in applied magnetic field~$B$ triggered
by QI can then be expressed as
\begin{equation}
\begin{multlined}
\label{eq:Hikami}
\updelta \sigma^\text{QI} (B) =
\frac{e^2}{\pi h}
\bigg[
\uppsi\left(\frac{1}{2}+\frac{\Bso+\Bphi}{B}\right)
-
\frac{1}{2}\uppsi\left(\frac{1}{2}+\frac{\Bphi}{B}\right)
\\
+
\frac{1}{2}\uppsi\left( \frac{1}{2} + \frac{2\Bso+\Bphi}{B}\right)
-
\uppsi\left(\frac{1}{2}+\frac{\Bo}{B}\right)
\bigg],
\end{multlined}
\end{equation}
with $\uppsi$ being the digamma function.\cite{dresselhaus:1992} The introduced effective magnetic
fields are related to the scattering times via
\begin{equation}
\Bo[/\phi/so] = \frac{\hbar}{4eD\tauo[/\phi/so]},
\end{equation}
with $D$ being the diffusion constant.

Magnetoconductivity in relevant magnetic fields $B\ll\Bo$ is then given by
\begin{equation}
\begin{multlined}
\label{eq:Hikami2}
\Delta\sigma^\text{QI} (B)
=
\updelta \sigma^\text{QI}(B) - \updelta \sigma^\text{QI}(0)
 \\
= \frac{e^2}{\pi h} \bigg[\Psi\left( \frac{B}{\Bso + \Bphi} \right)
-
\frac{1}{2} \Psi\left( \frac{B}{\Bphi} \right)
+
\frac{1}{2} \Psi\left( \frac{B}{2\Bso+\Bphi}\right)
\bigg],
\end{multlined}
\end{equation}
where $\Psi(x)=\ln(x)+\uppsi(\tfrac{1}{2}+\tfrac{1}{x})$.

In the \TwoD case, experimental data are often presented in terms of the related
MR calculated from the magnetic field-dependent resistance $R(B)$ via
\begin{equation}\label{eq:MR}
\MR = \frac{R(B)-R(0)}{R(0)}
= \frac{1}{1+\rho(0) \Delta\sigma^\mathrm{QI}(B)}-1,
\end{equation}
where the \TwoD resistivity $\rho$ is identified with the sheet
resistance $\Rs=\frac{w}{l}\cdot R$ with $l$ and $w$
being the measurement bar's length and width, respectively.

To compare the conductivity influenced either by magnetic fields or
temperature, Equation~\eqref{eq:Hikami} can be further adapted: Evaluating
$\updelta\sigma^\text{QI}(B)$ in the limit of zero magnetic field the first-order
correction to the conductivity can be individually expressed for both low-temperature
states associated with WL and WAL, respectively: in case of $\tauso\gg\tauphi$
($\Bso\ll\Bphi$) Equation~\eqref{eq:Hikami} treats WL and simplifies to
\begin{equation}\label{eq:Hikami_WL}
\updelta \sigma^\text{WL} (B \to 0) = \frac{e^2}{ \pi h} \ln \left( \frac{\Bphi}{\Bo} \right),
\end{equation}
whereas in case of $\tauso\ll\tauphi$ ($\Bso\gg\Bphi$) it relates to WAL and reads
\begin{equation}\label{eq:Hikami_WAL}
\updelta \sigma^\text{WAL} (B \to 0) =
-\frac{1}{2}\frac{e^2}{\pi h} \ln \left( \frac{\Bo^2 \Bphi}{2\Bso^3 } \right).
\end{equation}

\Bphi is controlled by inelastic scattering and an
algebraic temperature dependence of \Bphi is assumed by
\begin{equation} \label{eq:algebraic_dependence}
\Bphi(T) = \gamma + \beta T^\alpha,
\end{equation}
with $\beta$ being a scaling factor, $\gamma$ modeling a saturation in dephasing at zero
temperature, and $\alpha$ being an exponent in the range between 1 and 2
combining contributions of both electron–phonon and electron–electron
scattering.\cite{lin:2002,abrahams:1981}

With the help of Equation \eqref{eq:algebraic_dependence}, the first-order
quantum corrections to the conductivity become tempera\-ture-dependent with an
insulating state in case of WL
\begin{equation}
\updelta \sigma^\text{WL} (T) = \frac{e^2}{ \pi h} \ln \left( \frac{\gamma + \beta T^\alpha}{\CWL} \right),
\label{eq:sigmaTort}
\end{equation}
and with a metallic state in case of WAL
\begin{equation}\label{eq:Delta_sigma_T_WAL}
\updelta \sigma^\text{WAL} (T) = -\frac{1}{2} \frac{ e^2}{\pi h} \ln \left(
\frac{\gamma + \beta T^\alpha}{\CWAL} \right).
\end{equation}
Both progressions are exclusively driven by the temperature dependence of the
dephasing scattering with \CWL and \CWAL being temperature-independent
constants determined by WL and WAL, respectively.

An insulating ground state is not necessarily induced by Anderson localization but can
also be incited by EEI.\cite{lee:1985,altshuler:1980b,altshuler:1980a} In \TwoD
systems, the conductivity correction due to EEI reveals nearly the same logarithmic temperature dependence compared with WL
\begin{equation}\label{eq:Delta_sigma_T_EEI}
\updelta \sigma^\text{EEI}(T) = \frac{e^2}{\pi h} \ln \left( \frac{T^\zeta}{\CEEI} \right),
\end{equation}
with $\zeta$ being an exponent related to screening effects and ranging between
$0.35$ for no screening and 1 for perfect screening, and \CEEI being a
temperature-independent constant defined by EEI. The temperature
dependence can again be compared with magnetic field-dependent measurements as
in the presence of magnetic fields a finite Zeeman splitting (ZS) is
responsible for a sizable magnetoconductivity in \TwoD
systems:\cite{lee:1985}
\begin{equation}\label{eq:Delta_sigma_EEI_g2}
\Delta\sigma^\text{ZS}\big(\tilde B(T)\big)
=
-\frac{e^2}{\pi h}
\frac{2(1-\zeta)}{3} g_\mathrm{2D}\big(\tilde B(T)\big),
\end{equation}
with $\tilde B(T)=(g\muB B)/(\kB T)$, $g$ the Landé factor, and $g_\mathrm{2D}$ a function defined by
\begin{equation}
g_\mathrm{2D}\big(\tilde B(T)\big)
=\int\limits_{0}^{\infty}
\dd{\varOmega}
\ln\bigg\vert{1-\frac{\tilde B(T)^2}{\varOmega^2}}\bigg\vert\,
\frac{\mathrm d^2}{\mathrm d\varOmega^2}
\frac{\varOmega}{\exp(\varOmega)-1},
\end{equation}
which can be evaluated numerically.

\subsection{Sample Growth and Characterization of \BPO Thin-Films}

All samples discussed were grown by pulsed laser deposition (PLD). The PLD
system uses a KrF excimer laser with a wavelength of \SI{248}{nm} and a
nominal fluency of \SI{2}{\joule\per\square\centi\metre}. The used
polycrystalline \BPO targets were obtained commercially with asked maximum
achievable density. They are evaluated to have purities of at least
\SI{99.95}{\percent}. Prior to each sample growth the surface of the targets
were carefully cleaned.

\BPO thin-films were grown on commercially available, one-side polished,
(001)-oriented single-crystalline \STO substrates with a given size of
\SI{5 x 5 x 1}{\milli\metre}.
To obtain defined \BPO/\STO interfaces the substrates were either \TO
terminated using a hydrogen fluoride (HF) buffer
solution\cite{kawasaki:1994,koster:1998} and subsequently annealed in pure
oxygen flow at about \SI{950}{\celsius} for \SI{7}{\hour} or cleansed by lens
paper as well as ultrasonic bath treatment in acetone and isopropyl.

The substrates were then fixed for either infrared laser heating or resistive
heating on appropriate platforms using silver paste and transferred via a
load-lock system and transfer chamber into permanently air-sealed PLD
vacuum chambers. Depending on the pretreatment the substrates were either
slowly heated to nominally \SI{554}{\celsius} during at least
\SI{60}{\minute} in case of HF-treated substrates or heated up to
\SI{800}{\celsius} within a few minutes for at least
\SI{5}{\minute} in case of cleansed substrates to purify the substrate surface
and then reheated to about \SI{554}{\celsius} within seconds, both in a pure oxygen
background pressure of about \SI{1}{\milli\bar}.

Thin-film deposition was done using a nominal laser pulse energy of
\SI{550}{\milli\joule} and \SI{650}{\milli\joule}---depending on the used PLD
chamber---at a laser frequency of \SI{5}{\hertz}. The number of laser pulses
was chosen individually resulting in desired thin-film thicknesses. With this
setup, the growth rate of \BPO was determined to be about \SI{0.34}{\nano\metre}
per laser pulse.

After thin-film deposition, the vacuum chamber was immediately filled with pure
oxygen to at least \SI{400}{\milli\bar}, whereas the sample was cooled to
about \SI{400}{\celsius} within \SI{3}{\minute} and kept at that temperature for
additional \SI{17}{\minute} for annealing. Then the sample was allowed to
freely cool-down to room temperature before the chamber was evacuated again for
unloading the sample.

Film thicknesses were routinely obtained by X-ray reflectivity (XRR).
Conducted XRR fits resulted in averaged surface and interface roughness
better than \SI{0.6}{\nano\metre} and \SI{0.7}{\nano\metre}, respectively.
X-ray diffraction (XRD) measurements indicate that all epitaxial \BPO
layers are (001)-oriented.

All samples were patterned into four-probe and Hall bar layouts using a
standard photolithography system (mercury arc lamp) followed-up by ion-milling.
To minimize contact resistances gold was sputtered onto the contact pads. All
samples were electrically contacted using copper wires (\SI{0.1}{\milli\metre}
in diameter) soldered to the puck and glued via silver paste to the samples.

All electrical transport measurements were carried out using a commercial
14-T physical property measurement system (PPMS) with an electrical transport
option (ETO). The applied AC currents were in the range of
\SIrange{0.1}{1}{\micro\ampere} with typical frequencies from \SIrange{70}{128}{\hertz}.

\subsection{Experimental Results and Discussion}

In this article, we account for the EEI contribution intrinsically involved in
the \MR data. Assuming both WAL and EEI contributing equally via
Equation~\eqref{eq:Hikami2} and \eqref{eq:Delta_sigma_EEI_g2}, we
self-consistently evaluate \MR and \RsT
within the following iterative scheme:

We start by applying Equation~\eqref{eq:Hikami2} to our raw MR data and extract the
WAL contribution neglecting any EEI contribution during the first iteration. Subsequently, with the help of Equation~\eqref{eq:Delta_sigma_T_WAL}, we are
able to subtract the WAL contribution to reveal the pure temperature-dependent
sheet resistance due to EEI which then provides a value of the screening factor~$\zeta$.
By accounting for a pronounced Zeeman splitting the MR data can
now be reevaluated again allowing for a priorly hidden EEI contribution that is described by Equation~\eqref{eq:Delta_sigma_EEI_g2} with a presumed Landé factor $g=2$. We
carry out this procedure successively until the screening factor~$\zeta$ settles to a constant
value. To avoid oscillations which may prevent convergence---as $\zeta$ is
close and limited to 1---we average the obtained $\zeta$ values within the
last three iterations.

\begin{figure}
\centering
\includegraphics[width=\linewidth]{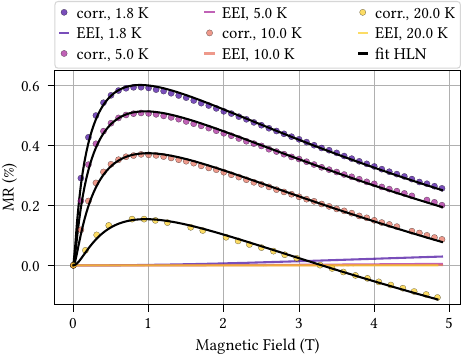}
\caption{%
\label{fig:MR}%
Reevaluated \MR data taken at different temperatures of a \tsampleone-thick \BPO
thin-film grown on a (001)-oriented single-crystalline \STO substrate. The
symmetrized raw data were self-consistently corrected from \EEI contributions
(Equation~\eqref{eq:Delta_sigma_EEI_g2}) with the presumed value of $g=2$ and
$\zeta=\zetasampleone$ retrieved from analysis of the \RsT measurement (see
Figure~\ref{fig:Delta_sigma_T}). The \EEI contribution for each temperature is
plotted as a solid line with its corresponding color. Black solid lines show
best fits (least squares method) of the \EEI-corrected MR data (see ``corr.'',
\ie, colored dots) using Equation~\eqref{eq:Hikami2} resulting in an averaged
value of $\Bso\approx\Bsosampleone$. The obtained temperature dependence of
\Bphi can be described by an algebraic dependence
(Equation~\eqref{eq:algebraic_dependence}) with $\alpha=\alphasampleone$,
$\beta=\betasampleone$, and $\gamma=\gammasampleone$ (not shown) determining the
WAL correction in the \RsT analysis (see Figure~\ref{fig:Delta_sigma_T}).}
\end{figure}

\begin{figure}
\centering
\includegraphics[width=\linewidth]{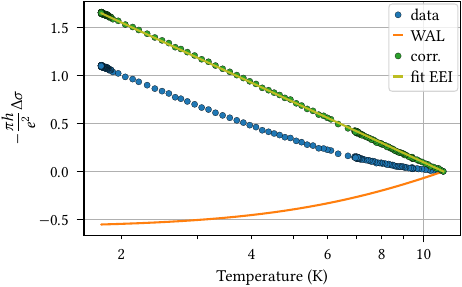}
\caption{\label{fig:Delta_sigma_T}%
Progression of the change in conductivity referenced to the temperature
$\Tref=\Trefsampleone$ in logarithmic scale. Blue dots show original measured
data, whereas the expected progression for WAL is plotted in orange, retrieved
from \MR analysis (see Figure~\ref{fig:MR}) following
Equation~\eqref{eq:Delta_sigma_T_WAL} indicating a metallic low-temperature state. In
green, WAL-revised data are shown which are perfectly explained by the \EEI
contribution solely (Equation~\eqref{eq:Delta_sigma_T_EEI})---resulting in
$\zeta=\zetasampleone.$}
\end{figure}

Exemplarily the result of such a self-consistent evaluation of MR and \RsT in
terms of WAL and EEI are shown in \textbf{Figure~\ref{fig:MR} and
\ref{fig:Delta_sigma_T}}. Figure~\ref{fig:MR} shows temperature-dependent
\MR data taken from a \tsampleone-thick \BPO thin-film showing an increase in
\MR to a maximum value at a magnetic field of $B\approx\SI{0.85}{T}$ with a
following decrease at higher magnetic fields, confirming our former
results.\cite{seiler:2019a} The MR data were corrected from concomitant EEI by
subtracting its contribution via Equation~\eqref{eq:Delta_sigma_EEI_g2}
with $\zeta=\zetasampleone$ retrieved from \RsT analysis. As expected, EEI
contributes only slightly (see colored lines in Figure~\ref{fig:MR}). The
reevaluated MR data can now be perfectly fitted in terms of WAL
using Equation~\eqref{eq:Hikami2}.

Further, the fits result in an averaged $\Bso\approx\Bsosampleone$ and a
temperature dependence of \Bphi that can be best described with
$\alpha=\alphasampleone$ following Equation~\eqref{eq:algebraic_dependence}
supporting a dephasing mechanism mainly due to electron–phonon scattering.

\begin{figure}
\centering
\includegraphics[width=\linewidth]{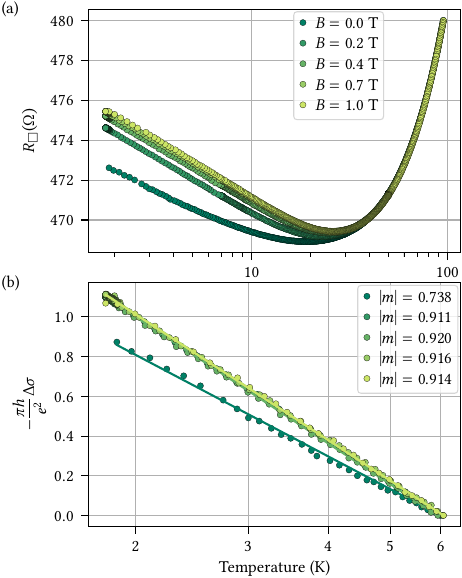}
\caption{\label{fig:Delta_sigma_T_B}
Progressions of \RsT as well as changes in conductivity normalized to
$\Tref=\SI{6}{K}$ while applying various perpendicular magnetic fields
between \SI{0.2}{\tesla} and \SI{1}{\tesla} (in logarithmic scale). The
magnetic field suppresses WAL contributions, whereas EEI contributions are
unaffected. The slopes clearly show trends toward an insulating ground
state as the magnetic field increases. The gradient $\vert m\vert$ was
linearly fitted, representing $\zeta$ in case of suppressed \WAL. The
resulting $\zeta$ ranging between \num{0.91} and \num{0.92} is in good
agreement with the prior self-consistent analysis ($\zeta=\zetasampleone$).}
\end{figure}

Simultaneously taken \RsT data are likewise affected by EEI at
low temperature, see zero-field data in \textbf{Figure~\ref{fig:Delta_sigma_T_B}}:
Upon cooling starting from room temperature \Rs steadily decreases, then reaches a
minimum at about \Trefsampleone and subsequently rises again. The
high-temperature progression can be well understood in terms of electron–phonon
scattering as well as thermally activated dislocation
scattering.\cite{fuchs:2015}

The low-temperature behavior is unequivocally controlled by quantum
corrections. Figure~\ref{fig:Delta_sigma_T} shows the change of the conductivity
\begin{equation}
\Delta\sigma=\left( \frac{1}{\rho(T)} - \frac{1}{\rho(\Tref)} \right)
\end{equation}
normalized to $\Tref=\Trefsampleone$. The measured data were reevaluated by
subtracting the influence of WAL following Equation~\eqref{eq:Delta_sigma_T_WAL}
with parameters acquired from evaluations of the \MR. The corrected data show a
clear logarithmic increase perfectly described by EEI following
Equation~\eqref{eq:Delta_sigma_T_EEI} that results in $\zeta=\zetasampleone$.

For consistency, we applied the just established self-con\-sis\-tent calculations of \Bso
and $\zeta$ to the data presented in\cite{seiler:2019a} comparing different
sample thicknesses: For the sample with thickness \SI{21.3}{nm}, the WAL
contribution expressed by \Bso changes in its average value from \Bsosampletwoold to \Bsosampletwo,
whereas EEI represented by $\zeta$ remains unchanged at a value of \zetasampletwo. The \SI{4.8}{\nano\metre}-thick sample shows a small increase
in \Bso from \Bsosamplethreeold to \Bsosamplethree in average, whereas $\zeta$ changes
from \zetasamplethreeold to \zetasamplethree. It will be interesting to
further study the thickness dependence on both the WAL and EEI contributions.

An independent approach to extract the EEI contribution without being affected
by WAL is the magnetic field dependence of \RsT. Magnetic fields $B>\Bphi$ cause the quantum corrections induced by QI ($\updelta\sigma^\QI_B$) to
become temperature independent\cite{carl:1989} and therefore to vanish by evaluating
\begin{equation}
\Delta\sigma^\QI_B(T)=\updelta\sigma^\QI_B(T)-\updelta\sigma^\QI_B(\Tref).
\end{equation}
Hence, in the presence of even small magnetic fields, the temperature dependence of the
conductance below \Tref should be solely reigned by EEI.

In Figure~\ref{fig:Delta_sigma_T_B}, the temperature-dependent progression of
\RsT as well as $\Delta\sigma(T)$ normalized to now $\Tref=\SI{6}{K}$ are
plotted, both in logarithmic scale. The magnetic field further increases \Rs
pronouncing the insulating ground state according to the expected suppression
of WAL effects. The gradient $\vert m \vert$ (which translates directly
into the value of~$\zeta$ in case of suppressed \WAL) extracted from linear
fits clearly increases and saturates at $\vert m \vert\approx\num{0.915}$
in reasonable good agreement with our previous result
($\zeta=\zetasampleone$).

\section{Magnetotransport in Multiband Systems in the Clean Limit}
\label{sec:theoretical-studies}

Magnetotransport studies have also been carried out on multiband oxide
heterostructures. For example, for the confined electronic system at the
interface of \LAO/\STO, an EEI contribution was suggested to dominate transport
at low temperatures.\cite{fuchs:2015} This interpretation was challenged in a
more recent WAL analysis within the framework of a semiclassical approach to
multiband magnetotransport.\cite{Seiler2018}

A fully quantum mechanical multiband treatment of WAL was established for
degenerate, isotropic t\textsubscript{2g} bands.\cite{Seiler2019a,Seiler2019b}
However, for various multiband systems, such as the electron system at the
\LAO/\STO interface, band hybridization at crossing points or rather lines is
present in the relevant filling regime. This so far has not been addressed
within a fully quantum mechanical approach to WAL.

Here, as a first step to a more realistic modeling, we develop a description of
magnetotransport in the presence of band crossings within an effective two-band
model for a defect-free lattice system. We investigate explicitly the Hall
conductivity in the presence of atomic and Rashba-like spin–orbit coupling.

Before we reexamine the prerequisites of magnetotransport of a single-band model
and the two-band case with its particular Hall conductivity, let us introduce the
generic model description.

We use a tight-binding representation for the Hamiltonian of a noninteracting
electron system in an infinite \TwoD crystalline lattice
\begin{equation}
\label{eq:hamiltonian}
\Hopt =
\sum_{j, l}\sum_{\mu, \nu}t_{\vec{R}_{j} -\vec{R}_l}^{\mu, \nu}
(\cc_{\vec R_{j}}^\mu)^\dag \cc_{\vec R_{l}}^\nu
=
\intBZ\dd[2]{\vec k}\sum_{\mu, \nu}h^{\mu,\nu}_{\vec{k}}
(\cc_{\vec{k}}^\mu)^\dag \cc_{\vec{k}}^\nu,
\end{equation}
where $\vec R_j$ is a lattice vector. Lowercase Greek letters $\mu, \nu$ label
the states within a unit cell. The integral over the lattice momenta $\vec k$ is
taken over the first Brillouin zone (BZ), the area of which we denote by
$\varOmega_\mathrm{BZ}$.

The coordinate operator is assumed to be diagonal in the chosen basis $\{\ket{\vec{R}_j, \mu}\}$
\begin{equation}
\label{eq:coordinate_operator}
\ropt =
\sum_j\sum_\mu (\vec R_j + \vec{d}^\mu)
(\cc_{\vec R_j}^\mu)^\dag \cc_{\vec R_j}^\mu
=
\intBZ \dd[2]{\vec k}\sum_{\mu}
(\cc_{\vec{k}}^\mu)^\dag\ci\grad_{\vec{k}}^{\vphantom{\mu}} \cc_{\vec{k}}^\mu,
\end{equation}
where $\vec{d}^\mu$ is the position of the state $\mu$ within the unit cell. For
the last equality to hold the Fourier transformation of the creation and
annihilation operators must be defined for each state $\mu$ individually with
respect to its exact position:
\begin{equation}
\label{eq:fourier_transform_ladder_operators}
\cc_{\vec{k}}^\mu = \frac{1}{\sqrt{\varOmega_\mathrm{BZ}}} \sum_{j}
\exp[-\ci \vec{k}(\vec{R}_j + \vec{d}^\mu)] \cc_{\vec R_j}^\mu.
\end{equation}

Linear response theory provides us with the Kubo formula for the electric
conductivity in the static limit
\begin{equation}
\begin{multlined}
\label{eq:kubo_conductivity_dc}
\sigma^{\alpha\beta}_\mathrm{DC}
=
-\ci\hbar \lim_{\eta\to 0^+}\int_{\varOmega_{\mathrm{BZ}}}
\frac{\dd[2]{\vec k}}{(2\pi)^2}\sum_{m, n}
\frac{f(E_{\vec{k}}^m)-f(E_{\vec{k}}^n)}{E_{\vec{k}}^m-E_{\vec{k}}^n}
 \\\times
\frac{
\langle{\vec{k}, m}|{\J^\alpha}|{\vec{k}, n}\rangle\langle{\vec{k}, n}|{\J^\beta}|{\vec{k}, m}\rangle
}{E_{\vec{k}}^m-E_{\vec{k}}^n
+ \ci\hbar\eta},
\end{multlined}
\end{equation}
where $\ket{\vec{k}, m}$ describes an eigenstate of the Hamiltonian in band $m$
and $E_{\vec{k}}^m$ the corresponding
eigenvalue.\cite{Kubo1957,Mahan2000,Allen2006} For numerical stability $\eta$
has to be kept finite, which may be roughly interpreted as a finite scattering
rate. The Fermi distribution $f(E_{\vec k}^m)$ actually also depends on the
chemical potential and temperature. The electric current operator $\Jopt$ in the
reciprocal basis can be written in terms of the gradient of the Hamiltonian
matrix~$h^{\mu,\nu}_{\vec{k}}$
\begin{equation}
\label{eq:current_operator_tb}
\Jopt
= -e\frac{\ci}{\hbar}[\Hopt, \ropt]
= -e\intBZ \dd[2]{\vec k}\sum_{\mu, \nu}
\left(\frac{1}{\hbar}\grad_{\vec{k}}^{\vphantom{\mu}}h^{\mu,\nu}_{\vec{k}}\right)
(\cc_{\vec{k}}^\mu)^\dag \cc_{\vec{k}}^\nu,
\end{equation}
where $e$ is the elementary electric
charge.\cite{Boykin1995,Tomczak2009,Boykin2010}

As the coordinate operator (\cref{eq:coordinate_operator}) is assumed to be
diagonal, the effect of a homogeneous external magnetic field on the orbital
degrees of freedom is given purely in terms of the Peierls
phase.\cite{Peierls1933,Luttinger1951} No further parameters enter the model
description.\cite{Foreman2002,Boykin2010a} In general, the Hamiltonian will then
not commute with the lattice translation operators $\T_{\vec{R}_j}$, because of
the real space dependence of the vector potential. For a homogeneous external
magnetic field with rational flux $p/q$ per \TwoD unit cell, in units of the
magnetic flux quantum $\varPhi_0=h/e$, translation symmetry can be restored by
introducing magnetic translation operators
$\T_{\vec{R}_j}^\mathrm{M}$.\cite{Zak1964,Zak1964a,Brown1964} Those are a
combination of a gauge transformation and a lattice translation. They do not
commute with each other except if transporting a particle to the opposite corner
of a parallelogram penetrated by an integer number of magnetic flux quanta. The
smallest such parallelogram with a nonvanishing area is the so-called magnetic
unit cell, which is a $q$ times enlarged version of the lattice unit cell, so
that it is penetrated by an integer number~$p$ of magnetic flux quanta. Here and
in the following $p$ and $q$ are assumed to be coprime integers.

The quantum numbers of the commuting magnetic translation operators are good
quantum numbers to characterize the eigenstates of the Hamiltonian. They replace
the lattice momenta of the translation invariant system, resulting again in a
Hamiltonian in reciprocal space of the form of \cref{eq:hamiltonian}, where
$\mu$, $\nu$ now label the states in a magnetic unit cell. From a band
perspective, the enlargement of the unit cell to a magnetic one leads to a
splitting of each of the initial dispersion relations without field into $q$
magnetic Bloch bands (so-called Hofstadter bands). Each of the Hofstadter bands
contains only a fraction $1/q$ of the states of the original
bands.\cite{Hofstadter1976}

Under applied magnetic field the matrix elements of the current operator in the
eigenbasis of the Hamiltonian, as appearing in \cref{eq:kubo_conductivity_dc},
have the same $q$-fold degeneracy in the magnetic BZ as the
eigenvalues.\cite{Hofstadter1976} The integral over $\vec{k}$ must therefore in
the magnetic case only be taken over a reduced part of the magnetic
BZ.\cite{Chang1996,Arai2011,Mugel2017}

\subsection{Anisotropic Hofstadter Model} \label{sec:hofstadter}

Within this framework, we now consider a square lattice with one orbital per site
and nearest-neighbor hopping only:
\begin{equation}
\label{eq:anisotropic_hofstadter_hamiltonian}
\Hopt
=
\intBZ\dd[2]{\vec k}\left[-2t_\mathrm{x}\cos(k_\mathrm{x}) - 2t_\mathrm{y}
\cos(k_\mathrm{y})\right]
\cc_{\vec{k}}^\dag \cc_{\vec{k}}^{\phantom{\dag}}.
\end{equation}
The lattice spacing is set to 1 and spin polarization is assumed. We note that a
rectangular lattice geometry would in the following only lead to a scaling of
longitudinal conductivities and densities of states. We allow for an asymmetry
in the hopping strength along the two different bond directions. By introducing
the Peierls phase to account for a homogeneous magnetic flux through the lattice
cells, one arrives at the Harper--Hofstadter
Hamiltonian.\cite{Harper1955,Hofstadter1976}

\begin{figure}
\centering%
{%
\includegraphics[width=\linewidth]{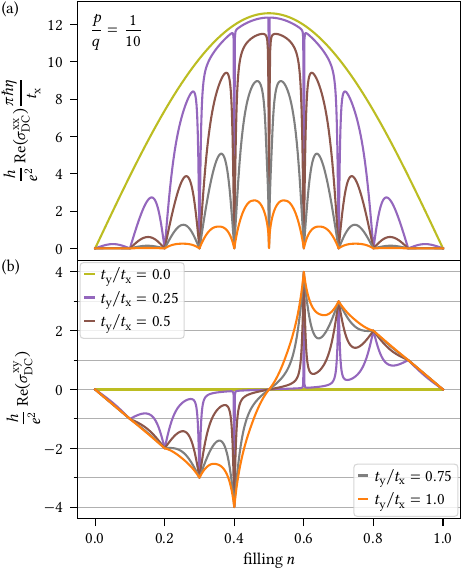}%
\phantomsubcaption{\label{fig:hofstadter_1_10:sigma_xx}}%
\phantomsubcaption{\label{fig:hofstadter_1_10:sigma_xy}}%
}%
\caption{\label{fig:hofstadter_1_10}
Quantization of conductivities: a)~longitudinal and b)~transversal
signals for different anisotropy values in the Hofstadter model
plotted versus band filling with a magnetic flux per unit cell of $p/q =
1/10$ of a magnetic flux quantum. The evaluations are done setting $k_\mathrm{B}T =
5t_\mathrm{x}\cdot10^{-4}$ and $\hbar\eta = t_\mathrm{x}\cdot10^{-3}$. The
periodicity of the field-free system is taken to be $N=12000$ lattice cells
in each direction.}
\end{figure}

To review how band structure and topology affect the conductivity of the
anisotropic Hofstadter model, we first choose a flux of $p/q = 1/10$. The
original cosine band is then split up into $q=10$ separate Hofstadter bands, as
long as the system is truly \TwoD ($t_\mathrm{x} \neq 0 \neq t_\mathrm{y}$). In
case of $q$ being even the two middle sub-bands in the Hofstadter model
touch.\cite{Hofstadter1976,Bernevig2013} All other bands are isolated by finite
energy gaps and have a Chern number of $+1$.\cite{Bernevig2013,Thouless1982}
This can be verified in \textbf{\cref{fig:hofstadter_1_10}}, as the longitudinal
conductivity vanishes in those gaps, whereas the transversal conductivity is
quantized in units of the conduction quantum $e^2/h$. This holds approximately
true even at finite temperatures and scattering rates, as long as temperature
$k_\mathrm{B}T$ and scattering-induced energy broadening $\hbar\eta$ are much
smaller than the bandgaps. On the other hand, if the chemical potential is
placed within a Hofstadter band, one calculates a finite Drude weight in case of
the longitudinal conductivity and the Hall signal is shifted away from its
quantized values.

In the limit of a 1D system with either $t_\mathrm{x}=0$
or $t_\mathrm{y}=0$, the Peierls phase can be gauged away completely. One is
left with the field-free model with a single band with zero Hall signature.

As the anisotropy between the hopping parameters in x- and y-directions is
in-/decreased, only the contributions to the conductivities, which are not of
topological character, approach the fully an-/isotropic limit (see yellow/orange
lines in \cref{fig:hofstadter_1_10}). For filling factors $n = r/q$, on the other
hand, where $r$ Hofstadter bands are completely filled, the conductivities are
invariant as long as no single energy gap becomes too small.

\begin{figure}
\centering%
{%
\includegraphics[width=\linewidth]{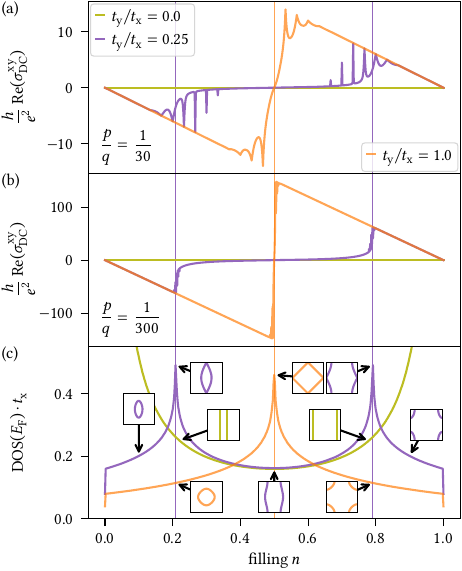}%
\phantomsubcaption{\label{fig:hofstadter_1_30:sigma_xy}}%
\phantomsubcaption{\label{fig:hofstadter_1_30:sigma_xy_300}}%
\phantomsubcaption{\label{fig:hofstadter_1_30:DOS}}%
}%
\caption{\label{fig:hofstadter_1_30}
Hall conductivity in respect to open and closed semiclassical orbits: Hall
signal for a)~$p/q=1/30$ of a flux quantum per unit cell and
b)~$p/q=1/300$. All other parameters are unchanged from
\cref{fig:hofstadter_1_10}. c)~Density of states of the field-free model at
the Fermi energy in dependence of the filling factor. The insets show the
Fermi surfaces at certain fillings in the first BZ, where the horizontal
axis represents $k_\mathrm{x}$ and the vertical axis $k_\mathrm{y}$. In all
three subfigures, the positions of the logarithmic Van Hove singularities of
the model with anisotropy $t_\mathrm{y}/t_\mathrm{x} = 0.25$ and the
isotropic case are indicated by vertical lines.}
\end{figure}

By reducing the magnetic flux (\textbf{\cref{fig:hofstadter_1_30}}) for a fixed
value of the anisotropy with $0 < t_\mathrm{y}/t_\mathrm{x} < 1$, one can see
that the Hall signal is filling-wise divided into distinct regimes where it
either approaches the fully anisotropic or the isotropic limit. The same holds
true also for the longitudinal conductivity. The boundaries between those
different cases are associated with the positions of the logarithmic Van Hove
singularities of the field-free model.\cite{Gobel2020} This is reasonable if one
recalls that those two Van Hove singularities originate from the saddle points
of the band structure and are thus at the same fillings as the transitions
between different kinds of semiclassical orbits.\cite{VanHove1953} In this
specific case, one finds closed orbits for low and high fillings of the
anisotropic Hofstadter model, whereas in between the logarithmic Van Hove
singularities only open orbits exist (see insets in
\cref{fig:hofstadter_1_30:DOS}). The isotropic limit is a special case: the two
considered Van Hove singularities merge in energy, which leads to an immediate
switching from electron to hole-like closed orbits, with only a single energy
level in between accommodating open orbits.\cite{Arai2010} In the fully
anisotropic limit, on the other hand, there are only open orbits, which are purely
1D and yield no Hall signal as already mentioned.

The sharp topological peaks in the regions of open orbits that one finds for
high magnetic fields (\cref{fig:hofstadter_1_10:sigma_xy}) are washed out
quickly with decreasing magnetic field by finite temperatures and scattering, as
there the gaps between the Hofstadter bands become small.

From semiclassical Boltzmann transport theory, one can deduce an expression for
the nontopological contributions to the Hall conductivity of the considered
band model at zero temperature, assuming $t_\mathrm{y} \leq t_\mathrm{x}$:
\begin{equation}
\label{eq:Hall_Boltzmann_theory}
\sigma^\mathrm{xy}_\mathrm{DC}
=
-\frac{e^2}{h}\frac{q}{p}
\left(n - \frac{|\overline{k}_\mathrm{x}(n)|}{\pi}\right),
\end{equation}
where $|\overline k_\mathrm{x}(n)|$ is the absolute value of the time averaged
$k_\mathrm{x}$-value along a semiclassical orbit at the Fermi surface for a
certain band filling $n$ (compare\cite{Azbel1987,Ashcroft1976}). So in case of
only closed electron orbits $|\overline{k}_\mathrm{x}| = 0$ and in case of
exclusively closed hole orbits at the Fermi level $|\overline{k}_\mathrm{x}| =
\pi$, whereas for open orbits, $|\overline{k}_\mathrm{x}(n)|$ is bounded by the
minimal and maximal absolute $k_\mathrm{x}$-value of the open orbit. Thus,
opposed to the standard textbook derivations, \cref{eq:Hall_Boltzmann_theory} is
not limited to the linear contributions of closed electron or hole orbits to the
transversal conductivity.\cite{Ashcroft1976} It describes the complete filling
range, even the suppression of the Hall signal for open orbits and the switching
from electron to hole-like behavior at half filling.

For a similar study about open and closed orbits in
the Hofstadter model where the anisotropy is due to a diatomic
basis see\cite{Gobel2020}.

\subsection{Effective Two-Band Model in a Perpendicular Magnetic Field}
\label{sec:twoBandModel}

With knowledge of the magnetotransport behavior of the single-band model from
\cref{sec:hofstadter}, one can now proceed to study a multiband system, where
two such square lattice cosine bands are combined. Its field-free Hamiltonian is
given by
\begin{equation}
\begin{multlined}
\label{eq:twoBandModel_Ham}
\Hopt = \intBZ\dd[2]{\vec k}\sum_{\mu=1,2}\Bigg\{
-2 [t_{\mathrm{x}}^\mu\cos(k_\mathrm{x})+
t_{\mathrm{y}}^\mu\cos(k_\mathrm{y})]
(\cc_{\vec{k}}^\mu)^\dag \cc_{\vec{k}}^{\mu}
 \\
+
\epsilon^\mu(\cc_{\vec{k}}^\mu)^\dag
\cc_{\vec{k}}^\mu
+
\sum_{\nu=1,2}
\varDelta(k_\mathrm{y})\tau_\mathrm{x}^{\mu,\nu}
(\cc_{\vec{k}}^\mu)^\dag \cc_{\vec{k}}^{\nu}\Bigg\},
\end{multlined}
\end{equation}
where $\epsilon^\mu$ allows for a relative energy shift of the two bands against
each other, $\tau_\mathrm{x}$ is the first Pauli matrix, and
$\Delta(k_\mathrm{y})$ controls a spin–orbit-like coupling effect (see the
following text). We assume that both states in a unit cell ($\mu = 1,2$) are
centered at the same point ($\vec{d}^1 = \vec{d}^2$).

To provide a specific example of a perovskite oxide,
Hamiltonian~\eqref{eq:twoBandModel_Ham} can accommodate each reduced set of two
out of the six spin–orbital states of the effective \LAO/\STO band
model.\cite{Khalsa2013,Zhong2013,Kim2013} As such, it allows us to study the
complex patterns of the Hall signal for every pair of bands individually,
without interference from a plethora of additional states. The interplay between
the anisotropic d\textsubscript{yz}-/d\textsubscript{zx}-bands of the 3d
t\textsubscript{2g} orbitals of titanium and the isotropic
d\textsubscript{xy}-band governs the main structure of the Hall signal of the
effective six-band model. From this perspective we now concentrate on the Hall
conductivity emerging from coupling of an anisotropic ($\mu=1$, $t_\mathrm{y}^1
= 0.25t_\mathrm{x}^1$) and an isotropic ($\mu=2$, $t_\mathrm{x}^2 =
t_\mathrm{y}^2 = t_\mathrm{x}^1$) cosine band.

Neglecting the energy shift in the effective \LAO/\STO band model due to spacial
anisotropy at the interface, these two bands are assumed to be aligned at their
bottom. This arrangement leads to a match in energy, and thus filling, of the
logarithmic Van Hove singularity of the isotropic band with the upper
singularity of the anisotropic band. A two-band model with slightly different
relative band positions would be treated analogously.

Two different coupling effects will be considered. A constant coupling term with
$\Delta(k_\mathrm{y}) = \gamma$ as it arises in the six-band model between the
d\textsubscript{xy}-band and the d\textsubscript{yz}-/d\textsubscript{zx}-bands
due to atomic spin–orbit coupling. Furthermore, a $\vec k$-dependent coupling
$\Delta(k_\mathrm{y}) = -\alpha\sin(k_\mathrm{y})$ is examined. It resembles
the coupling term between the d\textsubscript{xy}-band and the
d\textsubscript{yz}-/d\textsubscript{zx}-bands, introduced by the symmetry
breaking at the \LAO/\STO interface.\cite{Khalsa2013,Zhong2013}

\begin{figure}
\centering%
\includegraphics[width=\linewidth]{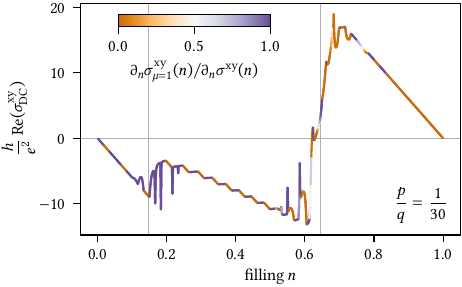}%
\caption{\label{fig:twoBandModel_relative_change}
Hall conductivity of an uncoupled two-band model ($\Delta(k_\mathrm{y})=0$)
broken down into distinct band contributions: anisotropic hopping
$t_\mathrm{y}^1 = 0.25t_\mathrm{x}^1$ in the first band ($\mu=1$) and
isotropic hopping $t_\mathrm{x}^2 = t_\mathrm{y}^2 = t_\mathrm{x}^1$ in the
second band. The two bands are aligned at their bottom
($\epsilon^\mu=(-1)^\mu 0.75 t_\mathrm{x}^1$). System size, temperature, and
scattering rate are chosen as in \cref{fig:hofstadter_1_10} and the magnetic
flux is at $p/q=1/30$. Vertical gray lines indicate the positions of
logarithmic Van Hove singularities. By calculating the relative change in
the Hall signal with filling, coming from state $\mu=1$ of the field-free
model, one obtains the impact of this state on the Hall conductivity at a
certain filling factor (see color coding). Loosely speaking, dark purple
sections result from the anisotropic band ($\mu=1$) and dark orange sections
are contributions from the isotropic band ($\mu=2$).}
\end{figure}

First, we inspect the Hall conductivity of the two uncoupled bands plotted
against the filling factor, as its structure already changes nontrivially with
respect to the single-band behavior studied in \cref{sec:hofstadter}. The
additional structural complexity is caused by the differing densities of states
of the two bands. Consequently, the conductivity of the uncoupled two-band
system may only be obtained by superposition of the individual signals after a
nontrivial transformation of each of them along the filling axis. By color
coding the total Hall conductivity
(\textbf{\cref{fig:twoBandModel_relative_change}}, purple sections belong to
$\mu =1$, orange sections refer to the orbital contribution $\mu =2$), the signal
is again resolvable from a single-band perspective.

In addition to the asymmetry of the signal with respect to half filling, which results
from the alignment of the two bands at their bottom, the most prominent new
feature in the Hall conductivity is a step-like descent for fillings between the
logarithmic Van Hove singularities. It should not be confused with the similar
looking quantized Hall conductivity resulting from gaps in the energy spectrum
when plotted against the chemical potential. In
\cref{fig:twoBandModel_relative_change}, the signal is shown versus band
filling, effectively skipping energy gaps in the dispersion specified by a
quantized Hall conductivity.

Thus, the ``treads'' of those steps cannot be the result of bandgaps. Instead,
they are the Hall signal of the wider Hofstadter bands of the anisotropic cosine
band, which has open semiclassical orbits in this range of filling, leading to a
nearly suppressed transversal conductivity.

The narrow energy gaps between those wider Hofstadter bands manifest themselves
in \cref{fig:twoBandModel_relative_change} as narrow ``topological peaks''
interrupting the horizontal progression of the step treads. However, as seen in
the single-band case in \cref{sec:hofstadter}, they are quickly washed out by
scattering and temperature, remaining only visible in the vicinity of the
logarithmic Van Hove singularities.

The step ``risers'', on the other hand, can be traced back to the flat
Hofstadter bands of the isotropic cosine band, corresponding to closed
semiclassical orbits. Typically, such a flat Hofstadter band (with $\mu=2$) is
placed energetically somewhere within a wider one (with $\mu=1$). When the
chemical potential reaches this flat Hofstadter band its much higher density of
states leads to a near total suspension of the filling up of the wider band,
until no empty states are left in the flat band. Thus, the slope of the Hall
conductivity changes abruptly compared with the step treads and the height of the
riser assumes a nearly quantized value (of $e^2/h$).

\begin{figure}
\centering%
{%
\includegraphics[width=\linewidth]{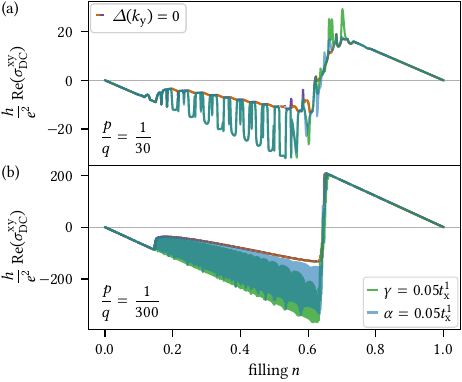}%
\phantomsubcaption{\label{fig:twoBandModel_weak_coupling:q30}}%
\phantomsubcaption{\label{fig:twoBandModel_weak_coupling:q300}}%
}%
\caption{\label{fig:twoBandModel_weak_coupling}
Evolution of the Hall conductivity for a hybridized two-band model: a)~Hall
signal from \cref{fig:twoBandModel_relative_change} ($\Delta(k_\mathrm{y}) = 0$) compared with the two
different types of band coupling $\Delta(k_\mathrm{y}) = \gamma$ and
$\Delta(k_\mathrm{y}) = -\alpha\sin(k_\mathrm{y})$ for small coupling
constants. b)~Same as a) but with a reduced magnetic field.}
\end{figure}

The regime with the step-like behavior is then expected to be heavily affected
already by adding a weak coupling term $\Delta(k_\mathrm{y})$ to the Hamiltonian
(\textbf{\cref{fig:twoBandModel_weak_coupling}}), as the different Hofstadter
bands will hybridize strongest at their intersection lines. In the case of a
weak magnetic field (\cref{fig:twoBandModel_weak_coupling:q300}), it is actually
the only range of filling where the Hall signal of the weakly coupled bands
differs significantly from the one of the uncoupled bands. It is striking that a
weak perturbation modifies the Hall signal qualitatively---an observation that
will be explained below. The other affected region around the coinciding
logarithmic Van Hove singularities (\cref{fig:twoBandModel_weak_coupling:q30}),
where the Hall signal switches its sign, will not be investigated closer, as it
shrinks to zero width in the low magnetic field limit.

\begin{figure}
\centering%
{%
\includegraphics[width=\linewidth]{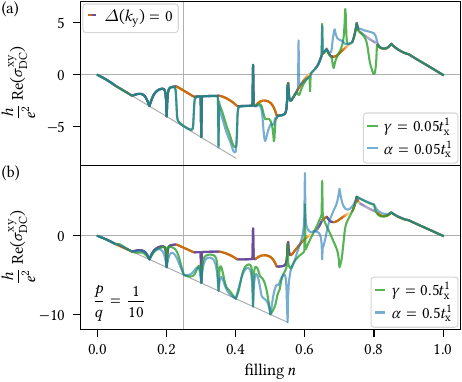}%
\phantomsubcaption{\label{fig:twoBandModel_coupling_p1q10:weak}}%
\phantomsubcaption{\label{fig:twoBandModel_coupling_p1q10:strong}}%
}%
\caption{\label{fig:twoBandModel_coupling_p1q10}
Reappearance of a topologically quantized Hall conductivity for hybridized
Hofstadter bands: Hall signal for the uncoupled model of
\cref{fig:twoBandModel_relative_change} ($\Delta(k_\mathrm{y}) =0$) with an
increased magnetic flux of $p/q=1/10$ compared with the two different types of
band coupling $\Delta(k_\mathrm{y}) = \gamma$ and $\Delta(k_\mathrm{y}) =
-\alpha\sin(k_\mathrm{y})$ for a) weak and b) strong coupling. The filling
of half an electron per unit cell ($n=0.25$) is marked by a vertical gray
line. For the lower fillings, the integer topological values of the Hall
signal line up (descending gray line), as there the magnetic Bloch bands all
have a Chern number of $+1$.}
\end{figure}

For weak coupling strengths, the deviation from the behavior of the uncoupled
bands in the step-like region can be well understood by first looking at higher
magnetic fields (\textbf{\cref{fig:twoBandModel_coupling_p1q10}}). Band
structure and Berry curvature are for weak coupling only distorted in the
vicinity of the former band crossings. So the Hall signal is expected to stay
mostly unchanged. It can only deviate significantly from that of uncoupled bands
in the filling ranges of the step risers (\eg, $0.2 < n < 0.3$, in
\cref{fig:twoBandModel_coupling_p1q10:weak}).

For a flat primary Hofstadter band intersecting a wider one, the shape of the
Hall signal of the hybridized bands can be constructed based on two facts: band
repulsion and the Chern numbers of the hybridized magnetic Bloch bands. By
hybridization, the wider primary Hofstadter band is split apart at the energy of
the flat band and each part is merged with half of the flat band, which is
itself split along the intersection lines. Thus forming two new nonintersecting
hybridized magnetic Bloch bands.

For weak coupling strengths, the new bands in the regions around the former
crossings are pushed above/below the energy of the primary flat Hofstadter band,
due to band repulsion. In contrast, in the other regions of the BZ, the band
dispersions and also the Berry curvatures are nearly unchanged. This means that
filling-wise the progression of the transversal conductivity only changes at the
two edges of the former step riser, whereas in the middle part of that region
one still finds the same linear trend as before.

For strong coupling, all hybridized magnetic Bloch bands in this regime are
energetically separated from each other by finite bandgaps. A Chern number of
$+1$ can in this case easily be read off from
\cref{fig:twoBandModel_coupling_p1q10:strong} for each of the new bands (see the
peaks at fillings of completely filled magnetic Bloch bands lined up along a
descending line). This must also hold true for the weak coupling case, assuming
the bands do not cross while reducing the coupling strength---albeit the
hybridized bands may eventually overlap if the flatter band has a finite width.

Somewhere in the middle of the former step riser the energetically lower one of
the two hybridized magnetic Bloch bands is completely filled. Assuming
energetically nonoverlapping bands or, equivalently, that the upper hybridized
band only contributes linearly up to this filling factor, the Hall signal must
thus already be shifted down to the descending gray line connecting the
integer topological values in \cref{fig:twoBandModel_coupling_p1q10:weak}.
Otherwise, the Chern numbers of the nonintersecting hybridized bands could not
be matched correctly. This leads to a broad dip replacing the step riser. It is
the separation of the bands due to the hybridization that causes this sizable
finite down shift of the Hall signal.

Inspecting the case of a slightly weaker magnetic field more thoroughly
(\cref{fig:twoBandModel_weak_coupling}), where the assumption of totally flat
primary Hofstadter bands is even more accurate, one sees that such a broad dip
appears at every former step riser. Thereby replacing the step-like descent by
an oscillatory behavior, varying between the signal of the uncoupled bands and
the ``topological limit''. The gaps between the wider Hofstadter bands,
associated with the anisotropic cosine band, must have also been slightly
enlarged by the band coupling. In particular one can now identify their narrow
peaks in the whole region between the logarithmic Van Hove singularities
(\cref{fig:twoBandModel_weak_coupling:q30}), where they were suppressed before
by finite temperature and scattering.

For higher temperatures, the energy broadening of $k_\mathrm{B} T$ will
eventually extend over the range of several magnetic Bloch bands. This then
leads to an averaging out of these oscillations. Lowering the magnetic field has
the same effect with the addition that new phenomena can arise due to a finite
coupling strength, which can then also depend on the specific form of
$\Delta(k_\mathrm{y})$.

\section{Conclusion}
We discussed \TwoD magnetotransport in the presence of spin–orbit
coupling in single-band systems with disorder as well as multiband systems in
the clean limit.

Experimentally, we extracted self-consistently both \WAL and
\EEI contributions emerging as first-order quantum corrections to the
electrical transport properties of thin \BPO films. Thus, we offer a consistent
way to interpret quantum corrections on \TwoD films to thoroughly
identify an electronically correlated and insulating low-temperature state.

Furthermore, going from a single-band system to a general multiband setup, we
investigated a defect-free lattice system which reveals a striking behavior when
electronic bands hybridize in the presence of a magnetic field. We first
reanalyzed the Hall conductivity of the anisotropic Hofstadter model, where open
semiclassical orbits lead to a deviation from the well-known linear behavior in
the electron density of closed orbits. This fundamental knowledge of the
single-band behavior of the conductivity then allowed us to fully understand an
uncoupled multiband system. The additional effects of a weak band coupling in
this multiband system can be explained by the hybridization of intersecting
Hofstadter bands instead of the field-free bands.

Hereafter, it would be intriguing to investigate a disordered system in a generic
multiband setup to merge the aspects investigated in our complementary
studies. The implementation of band hybridization into a generalized version of
the Iordanskii--Lyanda-Geller--Pikus theory will be challenging but allows for a
fundamental understanding of multiband quantum interference.

\begin{acknowledgments}
R.B. and E.L. contributed equally to this work.
Financial support by the Deutsche Forschungsgemeinschaft
(project number 107745057, TRR 80) is gratefully acknowledged.
\end{acknowledgments}

\newcommand{\noopsort}[1]{}
\end{document}